# Optimised Design of a Current Mirror in 150 nm GaAs Technology

Lua Ying Qian
Electrical and Electronics Engineering
Xiamen University Malaysia
Selangor, Malaysia
luayingqian@gmail.com

Chia Chao Kang
Electrical and Electronics Engineering
*Xiamen University Malaysia*
Selangor, Malaysia
chiachao.kang@xmu.edu.my

Tan Jian Ding
Electrical and Electronics Engineering
*Xiamen University Malaysia*
Selangor, Malaysia
jianding.tan@xmu.edu.my

Mohammad Arif Sobhan Bhuiyan*
Electrical and Electronics Engineering
Xiamen University Malaysia
Selangor, Malaysia
arifsobhan.bhuiyan@xmu.edu.my

Khairun Nisa Minhad
Electrical and Electronics Engineering
*Xiamen University Malaysia*
Selangor, Malaysia
khairunnisa.minhad@xmu.edu.my

Mahdi H. Miraz*
School of Computing and Data Science
*Xiamen University Malaysia*
Selangor, Malaysia
m.miraz@ieee.org

*Abstract*—**The Current Mirror (CM) is a basic building block commonly used in analogue and mixed-signal integrated circuits. Its significance lies in its ability to replicate and precisely regulate the current, making it crucial in various applications such as amplifiers, filters and data converters. Recently, there has been a growing need for smaller and more energy-efficient Radio Frequency (RF) devices due to the advancements in wireless communication, the Internet of Things (IoT) and portable electronics. This research aims to propose an improved and optimised CM design focusing on compactness and energy-efficient operation. Through a comprehensive methodology involving transistor sizing, biasing techniques, load resistance selection, frequency response stabilisation and noise analysis, the proposed high swing CM design achieves a gain of at least 6.005 dB, a reduced power consumption of 91.17 mW, a wide bandwidth of 22.60 kHz and improved linearity as well as accuracy through precise current matching and minimised mismatch. This optimised CM design will further boost the realisation of compact and lower power RF devices, contributing to the advancement of analogue circuit design techniques and enhancing system performance, accuracy and reliability.**

## I. INTRODUCTION

Relentless development in wireless communication has put pressure on refining and optimising radio frequency (RF) devices to further improve their performance matrix trade-off [1-2]. In general, current mirrors (CM) are required in these devices to demand highly-equal current replication and precise control to ensure that the components are truly performing their functions of mixing, amplification, signal conversion and frequency oscillation. Thus, a CM is rather critical as a basic block in maintaining operational integrity and efficiency in RF devices.

The basic structure of the CM is consisting of two or more transistors or MOSFETs connected in a specific configuration as illustrated in Fig. 1. The input current also known as the reference current, is applied into the gate terminal of the first transistor M1 (reference transistor) that acts as a reference current source (IREF) [3, 4]. As CM is a current regenerating circuit, the current through the input branch and the circuit will generate the same amount of current in the output branch [5].

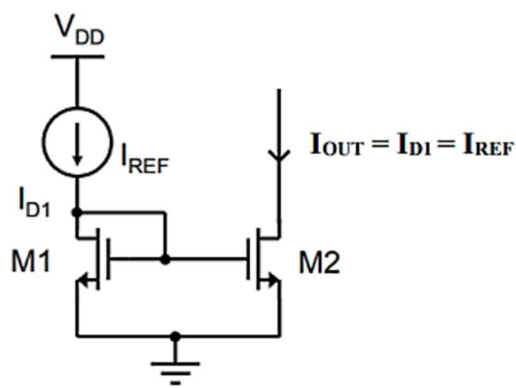

Fig. 1. The basic structure of the CM.

This structure is crucial as it serves as a current regenerating circuit, ensuring that the current through the input branch is reliably reproduced in the output branch. The effectiveness of a CM in an RF setting hinges on the exact matching of transistor parameters such as transconductance (gm) and threshold voltage (Vt). Adjustments in transistor sizes, biasing conditions and parameters are vital for achieving near-identical current replication, which is essential for the stability and performance of RF circuits.

Several researchers reported a number of diverse CM circuit architectures based on their operating principle which include basic CM (BJT CM), Wilson CM, Widlar CM and Cascode CM [5]. Depending on the requirements of CMs in RF devices, several factors such as process technology, power gain, linearity, bandwidth and noise performance should be taken into account. Various researchers proposed different circuit optimisation techniques to tune the most significant performance parameters for the desired applications. For example, a very low-power nonlinear nested CM design is reported in [6] which consumes only 0.000423 mW. The design by Liao *et al.* [7] achieve a remarkably low noise figure of 3.5 nV/√Hz that effectively minimises interference and signal degradation in RF devices. For enhanced bandwidth, Akbari *et al.* [8] reported another CM architecture based on flipped voltage-follower (FVF) for wideband RF devices and managed to achieve 2.383769 GHz bandwidth. Although the researchers adopted various optimisation techniques to improve the individual performance but achieving better overall performance trade-off is still the higher priority for portable RF devices.

In evaluating the performance of the current mirror (CM) for RF devices, both DC and AC analyses are essential to ensure proper functionality. DC analysis helps assessing the static behaviour of the circuit by verifying biasing, stability and load response, identifying potential issues such as thermal runaway or transistor saturation. AC analysis, on the other hand, provides information on the dynamic behaviour of the mirror circuit, which are essential for maintaining signal integrity in RF devices. Transistor aspect ratio optimisation and applying proper bias configuration can improvise the AC and DC analysis outcomes. If performance deviations are detected, the CM design is optimised by adjusting the capacitance, load resistance or biasing circuits [9, 10].

## II. CURRENT MIRROR DESIGN

In general, the CM design consists of four main blocks, viz. the reference current source, the current mirroring, the control and biasing as well as the output current mechanism, as shown in Fig. 2. The total design process can be divided into four steps, namely 1) CM configuration, 2) selection of transistors, 3) biasing as well as 4) current matching.

The configuration of the CM directly impacts the device performance, efficiency and stability. CM can be categorised into three main configurations: the basic CM, the Wilson CM and the high swing cascode CM. Each of these structures offers distinct advantages and challenges, making them suitable for different applications. The basic CM is simple but lacks precision, while the Wilson CM is more accurate but at the expense of increased circuit complexity. The high swing cascode CM has the best performance with a wide output voltage range but is more complex and costlier as well [11]. Although the choice depends on the specific requirements of the RF application, based on the flexibility of balancing factors, such as precision, voltage range, complexity and cost, the high swing cascode CM is the best choice [9]. Fig. 3 shows the schematic of the proposed high swing cascode CM circuit.

The selection of appropriate transistor is one of the most important aspects in designing CM circuit for RF devices. FETs are always preferred over bipolar junction transistors (BJT) due to power efficient and low noise operations. For this study, the UMS 150 nm pHEMT GaAs MOSFET was chosen because of its robust performance in the RF band. Its optimal pinch-off voltage, high drain-source leakage current and excellent transconductance portfolio make it an ideal candidate for ensuring efficient current regulation in RF devices. A simple CM with a cascode configuration, where common source (CS) and common gate (CG) stages act as input and output stages [12], respectively, can offer precise control over the biasing current. Such configuration can bring in better bias stability and current matching. A feedback resistor, connected between the input and output of the CM, acts as a sensing element to measure the output current and provides feedback to regulate the input current and thereby minimising the current mismatch errors between the reference and mirror path. Furthermore, an optimal load resistance has been utilised to ensure reasonable circuit gain and linearity.

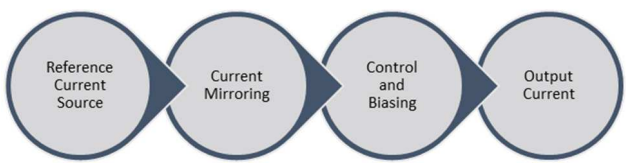

Fig. 2. The core blocks of a typical CM

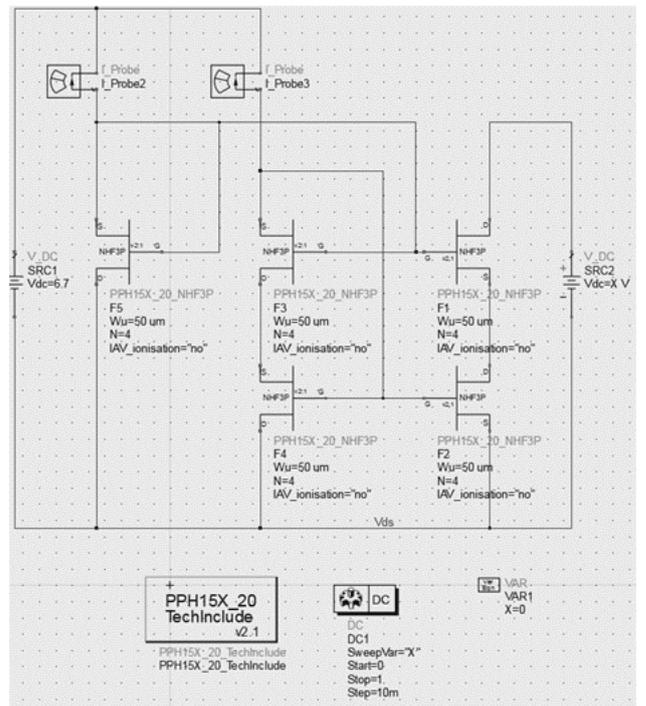

Fig. 3. The schematic of the proposed high swing cascode CM

## III. RESULTS AND DISCUSSIONS

The proposed high swing cascode current mirror circuit was designed and simulated using an UMS 150 nm pHEMT GaAs technology. The Electronic Design Automation (EDA) tools of Advanced Design System (ADS) software has been utilised for the simulation and layout design of the CM. In this paper, bias point analysis, direct current (DC) analysis and alternating current (AC) analysis have been carried out for performance evaluation.

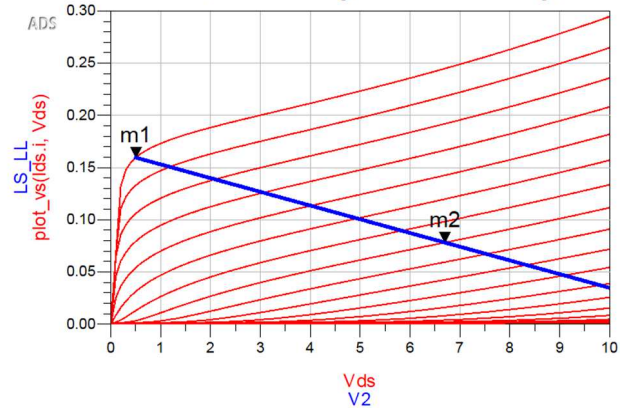

Fig. 4. The DC-IV simulation.

The bias point analysis plays a pivotal role in the configuration of the transistor, especially for RF applications. Proper biasing ensures that the transistor operates in the correct region of its characteristic curves and, therefore, will function with linear amplification of RF signals. The DC-IV simulation helps identify the appropriate gate voltage (Vgs) and drain current (Id) for the desired operating point.

For the appropriate transistor biasing, the DC-IV simulation results have guided us to select a biasing point for optimal performance, as illustrated in Fig. 4. The transistor is biased in such a way that Vds=6.7 V, while the gate is biased at -0.7 V. This is a precise biasing to keep the transistor in the most appropriate linear region for achieving amplification with the minimum possible distortion, necessary for RF applications. The Vds value of 6.7 V is within the safe operating area as outlined in the maximum ratings. This biasing point underpins the transistor's ability to deliver the best performance trade-off in this configuration. The flat region of the curve is the saturation region of the MOSFET, where it ensures constant amplification. Choosing the biasing point (m2) in Fig. 4 indicates that the transistor has been biased correctly for amplification in this CM circuit. This is the desired operating point for an RF amplifier, as it allows for a high gain while keeping the distortion to a minimum.

The DC simulation represents the basic analysis of the characterisation and design process of RF devices, since it allows the evaluation of the DC characteristics of the transistor. This analysis is able to select the appropriate configuration to achieve the most perfect current matching in a CM. The DC simulation specifically targets the study of the behaviour of the CM configuration for specific DC voltages. It allows the designers to tune the mirror path in such a way that the output current closely follows the input reference current, which is of paramount importance for the precision of the mirror and the overall performance in RF applications. The result from the DC simulation is shown in Fig. 5. This figure represents the effect of the variation of the secondary DC voltage sources (X) on the analysis of the current match, whose range is taken to be from 0.000 to 0.430. If X = 0.050V, the matching of the current is attained at -1.904A across the probe. Perfect current matching at the secondary voltage source (X) indicates that the circuit is well calibrated for an accurate replication of the current. This would infer that in analogue RF circuit applications, where exact matching of current is a prime factor, high swing cascode current mirror maintains accuracy in spite of the variability of the transistor characteristics.

The AC simulation is an important aspect in analysing the behaviour of RF circuits. It allows for the assessment of frequency response and gain behaviour, which are very important characteristics of every RF device. Fig. 6 illustrates logarithmic frequency scale, which is normally used for visualising the frequency response over a wide spectrum. From the graph, two important data have been extracted: the gain at low frequency (m1: 6.005 dB) and the -3 dB bandwidth (m2: 22.60 kHz). The gain of 6.005 dB represents the capability of the circuit to amplify the input signal at low frequencies, while the -3 dB bandwidth is the range of frequencies over which the circuit maintains a gain within 3dB of this maximum value. The total computed power consumption of the proposed high swing cascode CM circuit for its operation is approximately 91.17 mW.

| X | I_Probe2.i | I_Probe3.i |
|---|---|---|
| 0.000 | -1.910 A | -1.901 A |
| 0.010 | -1.908 A | -1.901 A |
| 0.020 | -1.907 A | -1.901 A |
| 0.030 | -1.905 A | -1.901 A |
| 0.040 | -1.903 A | -1.901 A |
| 0.050 | -1.901 A | -1.901 A |
| 0.060 | -1.900 A | -1.901 A |
| 0.070 | -1.898 A | -1.901 A |
| 0.080 | -1.896 A | -1.901 A |
| 0.090 | -1.895 A | -1.901 A |
| 0.100 | -1.893 A | -1.901 A |
| 0.110 | -1.891 A | -1.901 A |
| 0.120 | -1.890 A | -1.901 A |
| 0.130 | -1.888 A | -1.901 A |
| 0.140 | -1.886 A | -1.900 A |
| 0.150 | -1.885 A | -1.900 A |
| 0.160 | -1.883 A | -1.900 A |
| 0.170 | -1.881 A | -1.900 A |
| 0.180 | -1.879 A | -1.900 A |
| 0.190 | -1.878 A | -1.900 A |
| 0.200 | -1.876 A | -1.900 A |
| 0.210 | -1.874 A | -1.900 A |
| 0.220 | -1.873 A | -1.900 A |
| 0.230 | -1.871 A | -1.900 A |
| 0.240 | -1.869 A | -1.900 A |
| 0.250 | -1.868 A | -1.900 A |
| 0.260 | -1.866 A | -1.900 A |
| 0.270 | -1.864 A | -1.899 A |
| 0.280 | -1.862 A | -1.899 A |
| 0.290 | -1.861 A | -1.899 A |
| 0.300 | -1.859 A | -1.899 A |
| 0.310 | -1.857 A | -1.899 A |
| 0.320 | -1.856 A | -1.899 A |
| 0.330 | -1.854 A | -1.899 A |
| 0.340 | -1.852 A | -1.899 A |
| 0.350 | -1.851 A | -1.899 A |
| 0.360 | -1.849 A | -1.899 A |
| 0.370 | -1.847 A | -1.899 A |
| 0.380 | -1.846 A | -1.899 A |
| 0.390 | -1.844 A | -1.899 A |
| 0.400 | -1.842 A | -1.899 A |
| 0.410 | -1.840 A | -1.898 A |
| 0.420 | -1.839 A | -1.898 A |
| 0.430 | -1.837 A | -1.898 A |

Fig. 5. The current matching behaviour as a function of the external voltage.

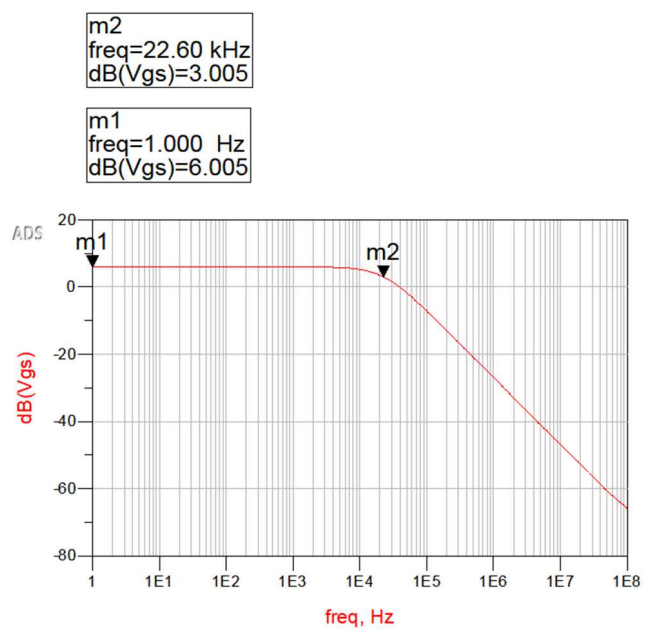

m2
freq=22.60 kHz
dB(Vgs)=3.005

m1
freq=1.000 Hz
dB(Vgs)=6.005

Fig. 6. The AC analysis graph.

In the comprehensive investigation of the high swing cascode current mirror, the experiment was carried out with great care, yielding robust data and insights. This performance evaluation of the proposed high swing cascode current mirror ranged from the characteristics of a transistor (FET) to detailed DC and AC simulations. The PPH15X 3P transistor traits were considered, outlining an in-depth focus on reducing its parasitic effects at 50 μm and 4-finger layout. The biasing point analysis confirmed the necessity for correct voltage and current settings, in order for the RF circuit to achieve optimal performance. Table I shows the summary of the performance of the proposed high swing cascode current mirror. The proposed design exhibits a reasonable trade-off amongst all the performance matrices.

TABLE I.  PERFORMANCE SUMMARY OF THE PROPOSED CM

| Parameters | Details |
|---|---|
| Chosen transistor | UMS 150-nm pHEMT GaAs process transistor |
| Configuration of current mirror | High swing cascode current mirror |
| Current accuracy | 100% when the external voltage source of 0.05V |
| Gain | 6.005 dB |
| -3dB Bandwidth | 22.60 kHz |
| Power consumption | 91.17mW |

## IV. CONCLUSION

In this paper, an implementation of the high-swing cascode current mirror with UMS 150 nm pHEMT GaAs process transistor for RF application has been reported. The proposed circuit had been characterised through four major parameters: accuracy, gain, bandwidth and power consumption. Our final results showed that the current mirror module had a current accuracy of 100% with an external voltage source of 0.05 V, a gain of 6.005 dB, a bandwidth of 22.60 kHz and a power consumption of 91.17 mW. Hence, this current mirror module is best suited for compact and low-power RF device applications.

## ACKNOWLEDGMENT

Xiamen University Malaysia has supported this research [Project code: XMUMRF/2021-C8/IECE/0021].